
\magnification=1200
\def\ni{\noindent}
\def\.{\mathaccent 95}
\def\beq{\begin{equation}}
\def\ee{\end{equation}}

\def\ep{\epsilon}

\def\la{\lambda}

\def\De{\Delta}

\def\frac#1#2{{\textstyle{{#1}\over {#2}}}}
\def\ni{\noindent}
\def\lsim{\mathrel{\rlap{\lower4pt\hbox{\hskip1pt$\sim$}}
    \raise1pt\hbox{$<$}}}
\def\gsim{\mathrel{\rlap{\lower4pt\hbox{\hskip1pt$\sim$}}
    \raise1pt\hbox{$>$}}}
\def\sqr#1#2{{\vcenter{\vbox{\hrule height.#2pt
         \hbox{\vrule width.#2pt height#1pt \kern#1pt
         \vrule width.#2pt}
         \hrule height.#2pt}}}}

% Next 5 lines define \lapprox and \gapprox: "less than or approximately
% equal to" and "greater than or approximately equal to".
\newbox\grsign \setbox\grsign=\hbox{$>$} \newdimen\grdimen \grdimen=\ht\grsign
\newbox\simlessbox \newbox\simgreatbox
\setbox\simgreatbox=\hbox{\raise.5ex\hbox{$>$}\llap
     {\lower.5ex\hbox{$\sim$}}}\ht1=\grdimen\dp1=0pt
\setbox\simlessbox=\hbox{\raise.5ex\hbox{$<$}\llap
     {\lower.5ex\hbox{$\sim$}}}\ht2=\grdimen\dp2=0pt

% 
% from Larry Molnar
% Set up some definitions:  
%
%This is how to have an approximate sign under < or > :

%\def\ref#1  {\noindent \hangindent=24.0pt \hangafter=1 {#1} \par}

\def\doublespace {\smallskipamount=6pt plus2pt minus2pt
                  \medskipamount=12pt plus4pt minus4pt
                  \bigskipamount=24pt plus8pt minus8pt
                  \normalbaselineskip=24pt plus0pt minus0pt
                  \normallineskip=2pt
                  \normallineskiplimit=0pt
                  \jot=6pt
                  {\def\smallskip {\vskip\smallskipamount}}
                  {\def\medskip   {\vskip\medskipamount}}
                  {\def\bigskip   {\vskip\bigskipamount}}
                  {\setbox\strutbox=\hbox{\vrule 
                    height17.0pt depth7.0pt width 0pt}}
                  \parskip 12.0pt
                  \normalbaselines}

                        %for boldface nabla
\font\gkvec=cmmib10                         %for boldface lowercase
                 %bold face nabla
    %bold face alpha
     %bold face beta
    %bold face gamma
    %bold face delta
        %bold face epsilon
     %bold face zeta
         %bold face eta
                  %bold face theta
     %bold face iota
    %bold face kappa
         %bold face lambda
       %bold face mu
       %bold face nu
                     %bold face xi
       %bold face pi
      %bold face rho
                  %bold face sigma
      %bold face tau
        %bold face upsilon
      %bold face phi
      %bold face chi
      %bold face psi
\def\bomega{\hbox{{\gkvec\char33}}}                  %bold face omega

\def\bw{\bar{\omega}}
\def\bv{\bar v}
\def\ts{\times}

\def\bfvp{{\bf v}'}
\def\bfjp{{\bf j}'}

\def\bfwp{{\bomega}'}
\def\bfbp{{\bf b}'}

\def\b0{b'^{(0)}}
\def\v0{v'^{(0)}}
\def\w0{\omega'^{(0)}}
\def\bb0{\bfbp^{(0)}}
\def\bv0{\bfvp^{(0)}}
\def\bw0{\bfwp^{(0)}}
\def\bj0{\bfjp^{(0)}}

\def\ni{\noindent}

\centerline{\bf Concave Accretion Discs and X-ray Reprocessing} 
\medskip
\centerline {Eric G. Blackman, Theoretical Astrophysics, Caltech 130-33, 
Pasadena CA, 91125, USA}
\centerline{(submitted to MNRAS)}

\centerline {\bf ABSTRACT}

Spectra of Seyfert Is are commonly modelled as emission 
from an X-ray illuminated flat accretion disc orbiting 
a central black hole. This provides both a reprocessed and direct
component of the X-ray emission as required by observations of
individual objects and possibly a fraction of the 
cosmological X-ray background.    
There is some observational motivation to at least consider the 
role that an effectively concave disc surface might play: 
(1) a reprocessed fraction $\gsim 1/2$ in some Seyferts
and possibly in the X-ray background, and 
(2) the commonality of a sharp iron line peak for Seyferts at 6.4KeV
despite a dependence of peak location on inclination angle
for flat disc models. Here it is shown that a 
concave disc may not only provide a 
larger total fraction of reprocessed photons, but can also 
reprocess a much larger fraction of photons in its outer regions 
when compared to a flat disc.  
This reduces the sensitivity of the 6.4KeV peak location to the inner disc 
inclination angle because the outer regions are less
affected by Doppler and gravitational effects.  If the X-ray source
is isotropic, the reprocessed fraction is directly determined by 
the concavity. If the X-ray source is anisotropic, the  
location of iron line peak can still be determined by concavity but
the total reflected fraction need not be as large as for 
the isotropic emitter case.  
The geometric calculations herein are applicable to 
general accretion disc systems illuminated from the center.


\vfill
\eject
\doublespace
\centerline{\bf 1. Introduction}

Accretion discs are the standard paradigm to
explain a wide variety of luminous galactic and extra-galactic
accreting sources.  Steady accretion requires dissipation of gravitational 
energy which produces the observed luminosity. 
Emission from X-ray binaries and 
active galactic nuclei (AGN) is thought to result from accretion onto 
a central massive black hole (e.g. Pringle 1981; Rees, 1984).  
The X-rays originate from the inner-most regions of the accretion flow
and probe the associated dynamics and geometry.

X-ray spectra of Seyfert I AGN have been modeled by a combination
of direct and reprocessed emission (see Mushotzsky et al. (1993) 
for a review).  The direct component
is from hot $10^9$K electrons of optical 
depth $\le 1$. The reprocessing component (e.g. Guilbert \& Rees 1988;
George \& Fabian 1991) 
%emission lines, absorption, and reprocessed continuum features. 
is composed of: 
1) a  Compton thick material in a moderate state of 
ionisation believed to be the thin accretion disc at $\le 10^5$K,which  produces iron fluorescence features and
2) possibly a Compton thin, highly ionised ``warm absorber'' 
which produces absorption features below 3 keV.
If the disc extends to the inner stable orbit, a broad 
gravitationally red-shifted 
 $\sim 6.4$ KeV iron K $\alpha$ fluorescence line can be produced.
Material further inside can also contribute to the 
line (Reynolds \& Begelman 1997) making a distinction
between Kerr or Schwarzchild holes difficult (though see Young et al. (1998)).
Generally, the iron line shape provides a diagnostic 
for strong gravity (Fabian et al., 1995; Tanaka et al 1996). 
Similar lines have also been seen in galactic black hole candidates.
(Fabian et al. 1989; Done et al. 1992; Ebisawa 1996).

The best studied (post ASCA) iron line is that of MCG-6-30-15 
(Tanaka et al 1995; Iwasawa et al. 1996).  
Its profile varies with 
the continuum, but has been successfully modeled by reflection off of a 
flat Keplerian disc inclined $\sim$ 30 degrees to the line of sight. 
However, ASCA has observed the $\sim 6.4$ KeV 
Iron lines in 22 Seyfert Is (Nandra et al. 1997)  
and as a population, they have a dispersion of only $\pm 3$ 
degrees around $29$ degrees in 
the predicted inclination angle when modelled with a flat disc.
The peak of the line always appears at or near the rest frame
$6.4$KeV. Because the peak location is sensitive to the disc
inclination angle in flat disc models (Laor 1991), 
the low dispersion is not expected for a distribution of randomly
oriented discs---even in the presence of a dusty torus at large radii
which narrows the angular range over which Seyfert Is, by class, 
are selected.  The narrow iron line core near 6.4 KeV 
correlates with the intensity of the continuum flux when
averaged over $\sim few \ts 10^4$ seconds (Iwasawa et al. 1996).  Perhaps this indicates a corresponding distance between the
reprocessing region and the direct X-ray source.

A second issue is the ratio of total reprocessed to direct emission.
In some Seyferts like MCG-6-30-15 this may slightly exceed 1 
(Lee et al. 1998; Guainazzi et al. 1999),
although for a population of 11 objects, the ratio seems
to hover around 1 (Matt 1998).
Some models of the cosmological X-ray background (c.f. Fabian 1992)
also suggest that the reprocessed X-ray component may 
exceed the direct component by a ratio $\gsim 5$.  
If true, this might be explained by flat disc geometries by employing 
an anisotropic direct X-ray source through the inverse Compton process,  
(Ghisellini et al 1990; Rogers 1991), direct acceleration of 
electrons toward the disc (Field \& Rogers 1993),  
source motion (Reynolds \& Fabian 1997),
or general relativistic (GR) effects (Martocchia \& Matt 1996).
But other possibilities, such as concavity, deserve investigation.
It important to note however, that recent models of the X-ray background
(Comastri et al. 1995) do not require the reprocessed to direct 
emission ratio much different from 1.  

A disc whose outer parts are thicker 
or concave with increasing height at larger radii may play
a role in explaining the ubiquity of the iron line peak at
6.4KeV even if the total reprocessed to direct emission ratio 
inferred for a given object is of order 1.  
The concavity allows a larger fraction of the reprocessed emission
to be reprocessed in the outer parts of the
disc, away from the influence of the doppler and gravitational effects.
The total reprocessed to direct fraction depends not only on the concavity
but also on how isotropic the X-ray source is.  
For an isotropic X-ray source, the total reprocessed fraction
will be large if concavity plays a strong role in determing
where the iron line peak is.  For an anistropic souce however,
the concavity may play a strong role in determining where the iron
line source is even if the total reprocessed fraction is modest.

Additional ``concavity'' beyond that of the simplest 
Shakura-Sunayev (Shakura 1973) discs can reflect a thickening 
disc with a flare due to the particular vertical structure and
temperature profile (e.g. Keynon \& Hartmann 1987 in the context
of stellar discs.) Also, azimuthally dependent concavity 
can result from warping which may be tidally or radiatively 
driven (Terquem \& Bernout 1993,1996; Pringle 1996,1997), 
or induced by a wind.  Alternatively, the discs may incur a transition
from a thick torus to a thin disc inward. 
Here I do not consider the dynamics in detail and just 
parameterize the concavity to highlight some simple effects
related to the above observations.  
Concavity was considered by Matt et al. (1991)
primarily for its effect on the iron line equivalent width, 
but this may not be its most important role.  
In section 2, I explicitly derive the ratio of reprocessed to 
direct emission as a function of the curvature.  
Splitting the reprocessed contribution into components
emanating from inside and outside a critical radius $r_c$,
I then derive the ratio of contributions to the iron line 
from the outer (taken to be the ``narrow component'' around 6.4KeV)
and inner regions (taken to be the ``broad component'').  The calculation 
results and line profile examples are discussed in section 3 and 
section 4 is the conclusion.

\centerline{\bf 2. Concavity and the Reprocessed Emission}

{\bf 2.1 Basic Considerations}

Here I take he direct X-ray source is taken to be an isotropic point emitter  
located above a Schwarzchild hole. (Later I will comment on
the possibility of an anisotropic source.)
The exact location of the X-ray source(s) is  
unknown.  To marginally avoid considering reflection from material 
free-falling inside of radii $r=6$ (Reynolds \& Begelman 1997), I 
take the source height to be $H_e= 10$, where 
the $r$ and $H_e$ are in gravitational 
units of $R_g\equiv GM/c^2$, and $M$ is the hole mass.
Incident hard X-rays impinge onto the 
accretion disc and are scattered off an optically thin outer layer (see
Matt et al. 1996). 
The X-ray photons also excite fluorescent photons from iron atoms.
Below, the $6.4$ KeV iron line flux is taken to be 
proportional to the incident flux. This assumes that the 
ionisation parameter is low enough over all $r\gsim 6$ to ensure cold
iron line emission (Matt et al. 1996).  
The calculations then reduce to the geometry of Fig 1.

{\bf 2.2 Total Reprocessed vs. Direct Flux Ratios for Concave discs} 

First consider the total reprocessed flux.
The ratio of the observed flux in the reprocessed component
to that in the direct component is the ratio of 
the number of photons which intercept the disc over the number 
of which escape directly. For the parameters chosen above, the GR
effects are sub-dominant in all regions of the disc and comprise
a maximum 30\% correction.
Note however, that GR leads to more flux impinging on the disc,  
so ignoring GR at first highlights a different
way to achieve a a high reprocessed fraction, and thus an overall
lower limit.

In the Euclidean regime, the flux ratio of reprocessed to direct
emission is given by the ratio
of solid angle subtended by the disc to that subtended by the free space to
the observer.  The azimuthal angle drops out by symmetry.
Take the disc's reprocessing layer to have a height 
 $$
h(r)=a(r/6)^b\ [{\rm for}\ a(r/6)^b <r]\ ;\ r\ [{\rm for}\ 
a(r/6)^b>r],   \eqno(1)
$$ 
where the curvature index $b$ implies concavity for $b>1$, 
and $a\equiv h(r_{in})/r_{in}$ at the inner disc radius $r_{in}=6$.
The two regimes of (1) are imposed to enforce $h\le r$ at all radii.
The dashed curves of Fig. 2 show the $b$ for which $h(r_{out})=r$ for different
outer radii $r_{out}$. The ratio of reprocessed to direct flux, using the
angles shown in Fig. 1, is then 
$$F_{rep}/F_{dir}={{\int_{\pi/2-Tan^{-1}[(H_e-h(r_{out}))/r_{out}]}
^{\pi-Tan^{-1}[r_{in}/H_e]}
Sin \theta d\theta}\over {\int_0^{\pi/2-Tan^{-1}[(H_e-h(r_{out}))/r_{out}]}Sin \theta d\theta}}.
\eqno(2)$$
For a strictly flat disc, the ratio is given by (2) with $h(r)=0$,
and then $F_{rep}/F_{dir}<1$.

%For a more strongly curved disc whose height at the outer
%radius satisfies $h(r_{out})> H_e$,
%the intensity ratio is given by 
%$$F_{rep}/F_{dir}={{\int^{Tan^{-1}[r_{out}/(h(r_{out})-H_e)]}_{\pi-Tan^{-1}[r_{in}/H_e]}
%Sin \theta d\theta}\over {\int^0_{Tan^{-1}[r_{out}/(h(r_{out})-H_e)]}_{\pi-Tan^{-1}[r_{in}/H_e]}
%Sin \theta d\theta}}.\eqno()$$

%$$F_{rep}/F_{dir}={{\int_{Tan^{-1}[r_{out}/(h(r_{out})-H_e)]}^{\pi-Tan^{-1}[r_{in}/H_e]}
%%Sin \theta d\theta}\over {\int^0_{Tan^{-1}[r_{out}/(h(r_{out})-H_e)]}_{\pi-Tan^{-1}[r_{in}/H_e]}
%Sin \theta d\theta}}.\eqno()$$

\centerline{\bf 2.3 Concavity and the Reprocessed Iron Line}

Consider two components to the reprocessed iron line, 
motivated partly by the approach employed for the 
best-studied MCG-6-30-15 example (e.g. Iwasawa 1996).  
Take the first component to peak at the rest frame frequency 
and the second to be the remaining broad line.
After choosing a core width, one can derive a corresponding critical
radius, $r_c$, outside of which all the narrow core emission emanates.  
The role of concavity becomes apparent by 
comparing the flux ratios from the two disc regions for
flat vs. curved discs.

%{\bf Deriving the critical radius $r_c$}

Iwasawa et al (1996) and Nandra et al. (1997) consider a narrow component 
width $\sim \pm 2.5$\% around the 6.4 KeV peak rest energy in MCG-6-30-15
and for the population of 22 objects respectively.
We can estimate $r_c$ by assuming that the spread 
corresponds to a Doppler width enveloping the largest and smallest 
frequencies at $r_c$.  
%The orientation
%of the disc with the line of sight will skew this line somewhat but
%that is of secondary importance for present purposes at the relevant radii.  
%it may be corrugated, in which case the core line could be
%rather symmetric. Regardless of symmetry, the relevant width is narrow.
The Doppler shift at large $r$ is 
$$
\nu/\nu_e\simeq (1\pm \De v/c),\eqno(3)
$$
where $\nu$ is the frequency,
$\nu_e$ is the rest frequency $=6.4$keV,  and
$\De v$ is the maximum spread in velocity.
For $\De v$ we take the Keplerian speed
$\sim c(1/r)^{1/2}$, 
%where $r_g\equiv(GM/c^2)$
%is the gravitational radius.  
Thus 
$$
r_c \simeq (\De\nu/\nu_e)^{-2},
\eqno(4)
$$
where $\De\nu$ is the frequency half-width of the core.
For $\De \nu=0.025$, $r_c=1600$,
while for  $\De \nu=0.05$, $r_c=400$.

%{\bf 3.3 Iron Line Flux ratios for flat vs. Concave discs} 

Using Fig. 1 to calculate the ratio of flux  emanating 
from outside $r_c$ to that from within $r_c$ we obtain 
$$F_{out}/F_{in}={{\int_{\pi/2-Tan^{-1}[(H_e-h(r_{out}))/r_{out}]}
^{\pi/2-Tan^{-1}[(H_e-h(r_c))/r_c]}
Sin \theta d\theta}
\over {\int_{\pi/2-Tan^{-1}[(H_e-h(r_c))/r_c]}
^{\pi-Tan^{-1}[r_{in}/H_e]}Sin \theta d\theta}}.
\eqno(5)$$
Notice from Fig. 1 that the angle bounds in the integral
are the same regardless of whether $h(r_c)>H_e$ or $h(r_c)<H_e$.
For a flat disc, (5) is found from taking $h=0$ for all $r$.   

{\bf 2.3 Local Approach}

The above approach is the simplest for the
Euclidean case, but to include GR and actually compute line profiles, 
GR corrections are more easily
incorporated in a formalism which integrates over $r$. 
An equivalent way to compute (5)
is to note that the reprocessed line flux from an element of area $dA$ 
is proportional to the impinging flux $f(r,H_e,h)$ at that radius projected
onto the area $dA$ (see Fig. 1)  
$$
dF\propto  g_r(r)f(r,H_e,h)Cos\la\ r(dr^2+dh^2)^{1/2}=
g_r(r)f(r,H_e,h)Cos\la(1+dh^2/dr^2)^{1/2}rdr$$
$$
\propto g_r(r)(r^2+(H_e-h)^2)^{-1}(1+dh^2/dr^2)^{1/2}rdr,\eqno(6)
$$
where 
$f(r,H_e,h)\propto (r^2+(H_e-h)^2)^{-1}$.
The angle $\la$ is between the normal to the area
element and the X-ray source direction (Fig. 1) so that 
$$
Cos\la=Cos(\pi/2-Tan^{-1}(dh/dr)+Tan^{-1}[(h(r)-H_e)/r]).\eqno(7)
$$
The quantity $g_r(r)=(1-2/r)^{-4}(1-2/r)^{1/2}(1+z)^{-4}$ 
is the product of three correction factors to an 
otherwise Euclidean formula: The first is the Doppler + GR
correction to the Euclidean illumination function as
approximated by Reynolds \& Begelman (1996).  The second
is the correction to the area measure in the integrand.
The third factor $(1+z)^{-4}$ is the red-shift correction.
This is given by $(1+z)=(p^{\mu}u_{\mu})_{em}/E_{obs}$
where subscript $em$ stands for emitted location (i.e. disc), $E_{obs}$ is the 
photon energy measured by the observer, 
and $p^\mu u_\mu$ is the
product of the photon 4-momentum and bulk motion 4-velocity along the sight line. At finite inclination angle, some of the photons
from the inner regions populate the narrow core.  
In addition, some inner region photons may be lost 
as they impinge on the outer disc and incur a second
reprocessing.  The formulae for a face on inner disc thus provide a lower 
limit to the ratio $F_{out}/F_{in}$ for all inclination angles. 
From  Reynolds \& Begelman (1996), we then obtain $(1+z)^{-4}=(1-3/r)^2.$  
Notice that the illumination function correction and the red-shift 
correction compete. The former enhances, while the latter decreases the 
flux.  The net effect is an increase
in the redshifted component and an increase flux from the inner regions,
and thus a decrease in (5).
I ignore the tiny corrections to $g_r(r)$ which result from a finite $h$.

%$$
%F_{in}/F_{out}={{\int_{r_{in}}^{r_c} g_r(r)  F(r',H_e,h)Cos\la r'dr'}
%\over {\int_{r_{c}}^{r_{out}} g_r(r) F(r',H_e,h)Cos\la r'dr'}}$$
%$$={{\int_{r_{in}}^{r_c} (H_e-h) (r'^2 +(H_e-h)^2)^{-3/2}r'dr'}
%\over {\int_{r_{c}}^{r_{out}} (H_e-h) (r'^2 +(H_e-h)^2)^{-3/2}r'dr'}}
%={{((H_e-h)^2+r_{in}^2)^{-1/2}-((H_e-h)^2+r_c^2)^{-1/2}}\over
%{((H_e-h)^2+r_{c}^2)^{-1/2}-((H_e-h)^2+r_{out}^2)^{-1/2}}}
%.\eqno(4)

Using (7) in (6) gives 
$$F_{out}/F_{in}=
{{[\int_{r_{c}}^{r_{out}} g_r(r')dr' r'(1+({dh\over {dr'}}
)^2)^{1/2}(r'^2+(H_e-h)^2)^{-1}
Cos\la(r')]
}\over{[\int_{r_{in}}^{r_c} g_r(r')dr' r'(1+({dh\over {dr'}})^2)^{1/2}(r'^2+(H_e-h)^2)^{-1}Cos\la(r')]}}.\eqno(8)$$
For a flat disc, $dh/dr=0$ and then $Cos\la=(H_e-h)/(r^2+(H_e-h)^2)^{1/2}$.
In the limit $g(r)=1$ for all $r$,  
the results from using (8) are {\it identical} to those from (5).  
In the next section I discuss the results of the above ratios
and show some line profiles.

\centerline{\bf 3. Results, Line Profiles, and Discussion}

The solid curves of Fig. 2 show $F_{rep}/F_{dir}$ as a function of 
$b$ for three values of $r_{out}$, using 
$a=1/50$, $H_e=10$, and $b\ge 1$.
%corresponding to $r_c=400$ and $r_c=1600$ respectively.  
The restriction to concave curvature (i.e. $b>1$) means that the entire 
disc surface sees the X-rays.  The imposed restriction 
that $h\le r$ leads to a maximum ratio of $5.4$ in Fig 2.
%The X-ray background requires a 
%ratio $\gsim 5$ (e.g. Fabian et al. 1990)
%so the effective concavity could play a role for large enough $b$.
For the above parameters, a Euclidean flat disc would 
give only $F_{rep}/F_{dir}\sim 0.87$ for an isotropic source.
Note that if the X-ray source were suitably anisotropic, 
the ratio $F_{rep}/F_{dir}$ need not be significantly 
greater than 1 even for the optimal $b$. In that case, the Euclidean
flat disc ratio for the same anisotropy would still be of order 5
times less for the optimal $b$.

Fig. 3 shows $F_{out}/F_{in}$,
for different values of $r_{out}$ and $r_c$.
(To be conservative, Fig. 3
employs (8) in order to include the rough GR corrections which  
$lower$ the curves relative to the Euclidean case, making
the effect $\sim 30\%$ 
less pronounced. By contrast, for Fig 2, the conservative 
lower limit is given by the Euclidean equation (2)).
There are two regions for each curve of  Fig. 3.
For low $b$, the gain from orientation toward the source wins 
over the decrease in flux from the extra distance to the disc at a given $r$
and the curves rise. However since $H_e<<r_c$, 
above the $b$ at which $h(r)/r=1$, 
the increasing distance between the X-ray point source 
and the disc for larger $r$ wins and the curves then decline. 
A Shakura-Sunaeyev (Shakura \& Sunaeyev 1973) 
disc corresponds to $b=9/8$.  For a flat disc the ratios
are low: for $r_c=400$, $F_{out}/F_{in}\sim 0.03$ and 
for $r_c=1600$, $F_{out}/F_{in}\sim 0.007$.

The $r_c=400$ solid curve lies completely above the 
$r_c=1600$ solid curve in Fig. 3 since $r_{out}$ is the same in both cases.
This contrasts the dashed curves which address a different issue.
The approximate $2-4\ts 10^4$ sec delay between an increase in continuum 
emission and the response 
of the narrow core line for e.g. MCG-6-30-15, (Iwasawa et al. 1996)
motivates considering that some reprocessing material resides at 
the associated distance from the X-ray source.  For a $10^7M_\odot$ hole, 
this corresponds to a distance of $400-800R_g$. The Keplerian velocities 
at this $r_c$ are consistent with a few percent frequency width of the core.  
This motivates determining the amount of reprocessed emission 
coming from $r_c \lsim r \lsim  10r_c$ and the results are 
the dashed curves of Fig 3. Because of the narrow range in $r$, the curves are 
lower than for the solid curves but there is still a range of $b$ for 
which such an outer ``ring'' can produce $F_{out}/F_{in}
\gsim 1/3$, even for $r_{out}=4000$, which is of order that 
required for MCG-6-30-15 (Iwasawa et al. 1996). 
By contrast, for flat discs from (2), even the more favorable case 
($r_{out}=10r_c$, $r_c=400$) gives only $F_{out}/F_{in}=0.03$.
The dashed curves in Fig. 3 cross because a decade in $r_c$ for higher 
$r_c$ is larger, but farther out, than a decade in $r_c$ for lower $r_c$.

The fact that emission in the narrow peak could
originate from large radii means that the ubiquity of observed
peaks near $6.4$keV in the 22 Seyferts studied by Nandra et al (1997)
would not be as sensitive to the inner disc inclination angle.
The maximum velocity dispersions at such large radii are small 
(e.g. $\De \nu\le 5\%$ at $r=400$). 
For a strictly flat disc, some tuning in inclination angle
is required to produce the location of the observed peak 
because gravitational+transverse
Doppler red, and blue shifts conspire to produce the particular peak values.
Figure 4 shows some line profiles using the
formulae of Fabian et al. (1989) (Esin 1998)
but with a modified emissivity function to include the flat and concave
disc cases. The emissivity function  of Fabian et al. (1989) was 
$\ep \propto (r_{in}/r)^2$ and I replace this by the factor
in (8): $(1+({dh\over {dr}})^2)^{1/2}(r^2+(H_e-h)^2)^{-1}Cos\la(r)$.
This simplifies for a flat disc as discussed below (8).
As expected from Figs. 2 and 3, Fig 4 shows the
narrow peak at 6.4KeV for the value of $b=1.7$
even at an inclination angle of $40$deg.  
The effect weakens significantly for $b=1.3$ as expected. 
Secondary reprocessing (Matt et al. 1991) of inner disc radiation by outer
disc radiation is not considered. This would be less likely
to effect the peak because the peak is produced from emission
at large radii. Also, the constraint
$h\le r$ means angles of inclination $< 45$deg are less 
affected by shadowing. 

The above results show that for a range of $b$ and $r_{out}$,
the ratios $F_{rep}/F_{dir}$ and $F_{out}/F_{in}$ can be
more than an order of magnitude larger than for flat discs.
The concavity may thus predominantly affect 
the total reprocessed fraction or the line shape rather than the 
line equivalent width (Matt et al. 1991).
The results are not strongly sensitive to the
parameters $a$, $H_e$, or $r_{in}$ for $H_e,r_e \gsim 6$,
but for $r_{in}, H_e<6$ a more detailed inclusion of 
the ionisation fraction and shadowing is required.
%That flat disc models have most all of the line core 
%emission coming from below $r_c$ by adjusting the disc 
%inclination angle means that (5) and (8) are lower limits:
%inner disc regions also contribute to the peak in figure 4.
%A high ionisation parameter in the inner regions
%also increases the ratio of $F_{out}/F_{in}$.
%The outer disc may also absorb some of the reprocessed emission from 
%the inner disc further amplifying (5) and (8).  

%(The authors come to this conclusion by investigating the improvement 
%in the fitting  when the exponent in the emissivity profile is allowed
%to be a free parameter)} ....
%Even for $r_{c}=1600$, corresponding to $\de \nu=2.5 \%$, a significant
%enhanced contribution to the line core comes from the outer
%regions when compared to that of a flat disc. 

The value of $b$ which provides both the largest
$F_{out}/F_{in}$ and $F_{rep}/F_{dir}$  
for the range of $r_{out}$ considered is $1.5\lsim b \lsim 1.9$
out to the radius for which $h(r)/r=1$.
(The value $b\sim 1.5$ is that of an isothermal disc.)
A disc which changes from a thin to thick disc/torus at $r\gsim r_c$
and has reprocessng material in the torus,  
may be approximated by (1).  
Some excess in reprocessed emission can also result from a warped
disc, possibly tidally or radiatively driven (e.g. 
Terquem \& Bertout 1993,1996; Pringle 1996,1997) or induced by wind torques,
but the azimuthal dependence must then 
be considered (e.g. Terquem \& Bertout 1993).
The role of an anisotropic X-ray source would change the total reprocessed
fraction, but not the influence on the line peak location,
or the comparisons to a flat disc.

\centerline {\bf 4. Conclusions}

Some simple but pronounced effects of a concavely curved accretion disc 
are captured by estimating the total reprocessed vs.
direct flux and the relative flux emanating
from inside and outside of a critical radius $r_c$.
For a range of $b$,  the reprocessed flux from $r>400$
and even $r>1600$ can contribute significantly to a rest frame iron line 
peak even when the (X-ray) radiation source is at located at $H_e\sim 10$.  
This alleviates some sensitivity of  
disc model predictions (e.g. Laor 1991) to the 
disc inclination angle, as seen in Fig 4.
Reprocessing at large distances also predicts a time 
delay between 
changes in the direct continuum emission and the reprocessed emission which may
be observed in some Seyferts 
(Iwasawa et al. 1996).  

In addition, concave discs may produce 
$F_{rep}/F_{dir}\gsim 5$ for an isotropic X-ray source
which provides another means for some AGN models of the X-ray background to
account for the high required reprocessed fraction (c.f. Fabian 1992). 
However, if the X-ray backround and Seyferts do not generically require
a high reprocessed fraction (Comastri et al 1995; Matt 1998),
the ubiquity of the iron line peak at 6.4 KeV could still be influenced
by concavity when compared to a flat disc:
The concavity affects not only the total reprocessed fraction for a given
X-ray source anisotropy, but the relative reprocessed fraction 
from different parts of the disc. 
Finally, 
note that the disc could be curved or could change from a thin to a thick 
disc with entrained reprocessing material, such that the effective global 
reprocessing geometry is approximated by a concavity.  

Acknowledgements: Many thanks to E. Chiang, C. Peres, 
and S. Phinney for discussions, to A. Esin for a subroutine, 
and to the referee for comments.

\ni Comastri A., Setti G., Zamorani G. Hasinger G., 1995, A\&A, 296, 1

\ni Dermer C.D., 1986, ApJ, 307, 47

\ni Dermer C.D., Liang E.~P., Canfield E., 1991, ApJ, 369, 410

\ni Done C. Mulcahaey J.S., Mushotzky R.F., Arnaud K., ApJ 395, 275. 

%\ni Celotti A., Fabian A.C., Rees, M.J., 1992, MNRAS, 255, 415.
 
\ni Ebisawa K., 1996, in {\sl X-ray Imaging and Spectroscopy of Hot Plasmas}, 
ed. F.Makino \& K. Matsuda, (Tokyo: Universal Academy Press) p427.
\ni Fabian A.C., Rees, M.J., Stella L., White N.E., 1989,
MNRAS, 238, 729.

\ni Esin A., personal communication.

\ni Fabian A.C., et al., 1995, MNRAS, 277, L11.

\ni Fabian, A.C., 1992, in {\sl The X-ray Background}
Barcons, X. and Fabian, A.C. eds., (Cambridge:  Cambridge Univ. Press), p305.

\ni Fabian A.C., George I.M., Miyoshi, S., Rees, M.J., 1990, 
MNRAS, 240, 14P.

\ni Field G.B. \& Rogers, R.D., 1993, ApJ {403}, 94.

\ni Galeev A.A., Rosner R., Vaiana G.S., 1979, ApJ, 229, 318

\ni Ghisellini G., George I.M., Fabian A.C., Done C., MNRAS, 1990, 248, 14.

\ni Guilbert P.W. \& Rees. M.J., 1988, MNRAS 233, 475.

\ni Guainazzi M. et al., 1999, A\&A, 341, L27.

\ni Haardt F., Maraschi L., 1993, 413, 507

%\ni Ichimaru S., ApJ, 1977, 214 840.

\ni Iwasawa K. et al., 1996, MNRAS, 282, 1038

\ni Kenyon S.J. \& Hartmann L., 1987, ApJ, 323, 714.

%\ni Kuncic, Z., Blackman, E.G., Rees, M.J, 1996, MNRAS 283, 1322.

\ni Laor A., 1991, ApJ, 376, 90

\ni Livio M. \& Pringle, J.E, 1996, 278, L35.

\ni Lee J.C., Fabian, A.C., Reynolds, C.S., Iwasawa, K. \& Brandt, W.N.,
1998, MNRAS, 300, 583.

\ni Martocchia A., Matt G., 1996, MNRAS, 282, L53.

\ni Matt G., 1998, astro-ph/9811053.

\ni Matt G., Fabian A.C., Ross R.R, 1996, MNRAS, 278, 111.

\ni Matt G., Perola G.C., Piro, L., 1991, A\& A, 247, 25.

\ni Mushotzky R.F., Done C.; Pounds K.A., 1997, ARA\&A, 31, 717

\ni Nandra K., George I.M., Mushotzky R.F., Turner T.J., Yaqoob T., 
1993, ApJ, 477, 602

%\ni Narayan R. \& Yi I., 1995, ApJ, 452, 710

%\ni  Narayan, R., Mahadevan, R. \& Quataert E., 1998a, 
%in {\it Theory of Black Hole Accretion} M.A. Abramowicz, G. Bjornsson, \& J.E. Pringle eds. (Cambridge: Cambridge Univ. Press).

%\ni  Pariev V.I. \& Bromley, B.C., 1998, astroph/9806134.

\ni Pringle J.E., 1996, MNRAS, 281, 357.

\ni Pringle J.E., 1997, MNRAS, 292, 136.

%\ni  Rees M.J.,Begelman M.C., Blandford R.D., Phinney E.S., 1982, Nature, 295, 17.

\ni  Rees M.J., 1984, ARAA, 22, 471.

%\ni  Rybicki G.~B., Lightman A.~P., 1979, {\it Radiative Processes in Astrophysics}, Wiley, NY

\ni Reynolds, C.S. \& Begelman, M.C., 1997, ApJ, 488, 109.

\ni Reynolds, C.S. \& Fabian, A.C., 1997, MNRAS, 290, L1.

\ni Rogers, R.D. \& Field G.B., 1991, ApJ, 370, L57.

\ni Rogers R.D. 1991, ApJ, 383, 550.

%\bibitem{} Shapiro S.~L., Lightman A.~P., Eardley D.~M., 1976, ApJ, 204, 187 

\ni Shakura, N.I. \& Sunyaev R.A., 1973, A\& A, 24, 337.

\ni Tanaka, Y. et al., 1995, Nature 375, 659.

\ni Terquem C. \& Bertout B., 1993, A\&A, 274, 291.

\ni Terquem C. \& Bertout B., 1996, MNRAS, 279, 415

\ni Young, A.J., Ross, R.R. \& Fabian, A.C., 1998, MNRAS, 300, 11.

\ni Zdziarski A.A., Fabian A.C., Nandra K., Celotti A., Rees
M.J., Done C., Coppi P.S., Madejski G.M., 1994, MNRAS, 269, L55

\noindent Zdziarski A.A., Johnson W.N., Done C., Smith D., McNaron-Brown K., 1995, ApJ, 438, L63

\vfill
\eject

{\bf Figure 1 Caption}: 
Schematic of X-ray source (small circle) above disc surface 
(thick black curve). Two values of $\theta(r)$ and $\la(r)$ angles 
are shown for illustration as used in (2), (5) and (8).
Regime (a) $H_e< h$, and  (b) $H_e >h$.

\medskip

{\bf Figure 2 Caption}: 
Plot of $h(r)/r$ (dashed curves) 
and $F_{rep}/F_{dir}$ (solid curves) vs. $b$ for 
$r_{out}=10^5,\ 16000,\ 4000$ from left to right in each curve group.
Values of $H_e=10$, $r_{in}=6$, and $a=1/50$ are used in all curves.
The flattening is the result of imposing $h(r)\le r$ for all $r$ in (1).
The X-ray source is assumed to be isotropic for this graph.
Simple anisotropies could be included by multiplying the y-axis
by a constant fraction.

\medskip

{\bf Figure 3 Caption}: 
$F_{out}/F_{in}$ vs. $b$ from (8)
for $r_{out}=10^5$ (solid curves) and $r_{out}=10 r_c$ (dashed curves).
The top solid and dashed curves as measured
at $b=1$ have $r_c=400$, and the bottom curves have $r_c=1600$.
The dashed curves cross and merge with the solid curves
above the $b$ for which $h(r)/r=1$.  The condition $h(r)\le r$ 
is imposed, but the presence of the down-turns only requires $r_c>>r_{in}$.

\medskip

{\bf Figure 4 Caption}: Line profiles for flat and concave models based on
Fabian et al. (1989) with $r_{out}=10^4$ with emissivity
function modified as discussed in text.  Solid lines 
are for flat discs at inclination angles of 
40 deg (broader curve) and 30 deg (narrower curve) respectively.
The concave disc curves match onto the wings of the flat curves.  
There are 4 curved disc profiles signatured by the height of their peaks.
The largest has $b=1.7$ and $30$ deg inclination.
The next largest is $b=1.7$ and $40$ deg. The
next is $b=1.3$ and $30$ deg, and the lowest is
$b=1.3$ and $40$ deg.

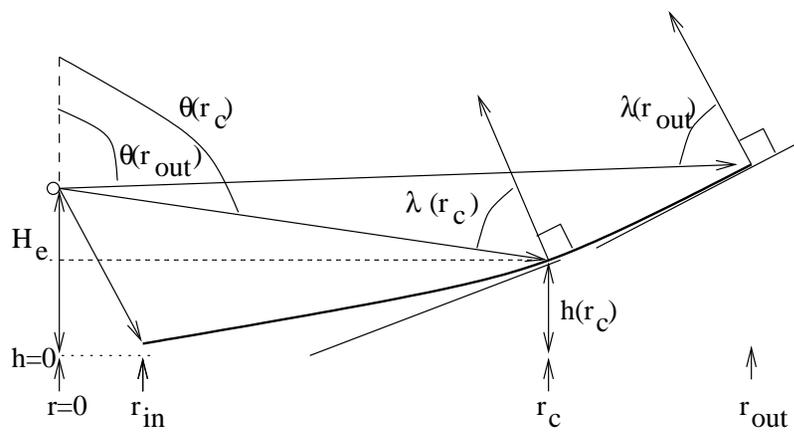
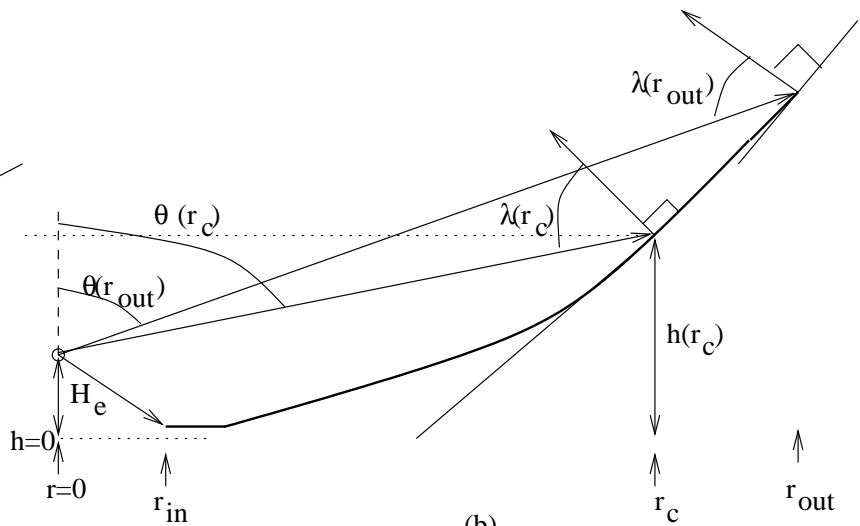
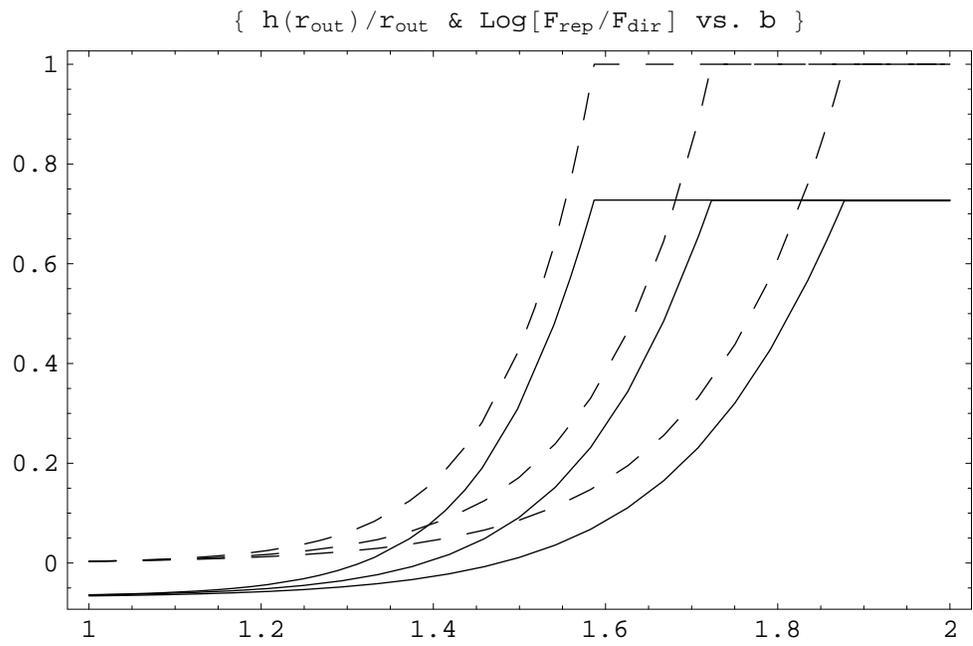
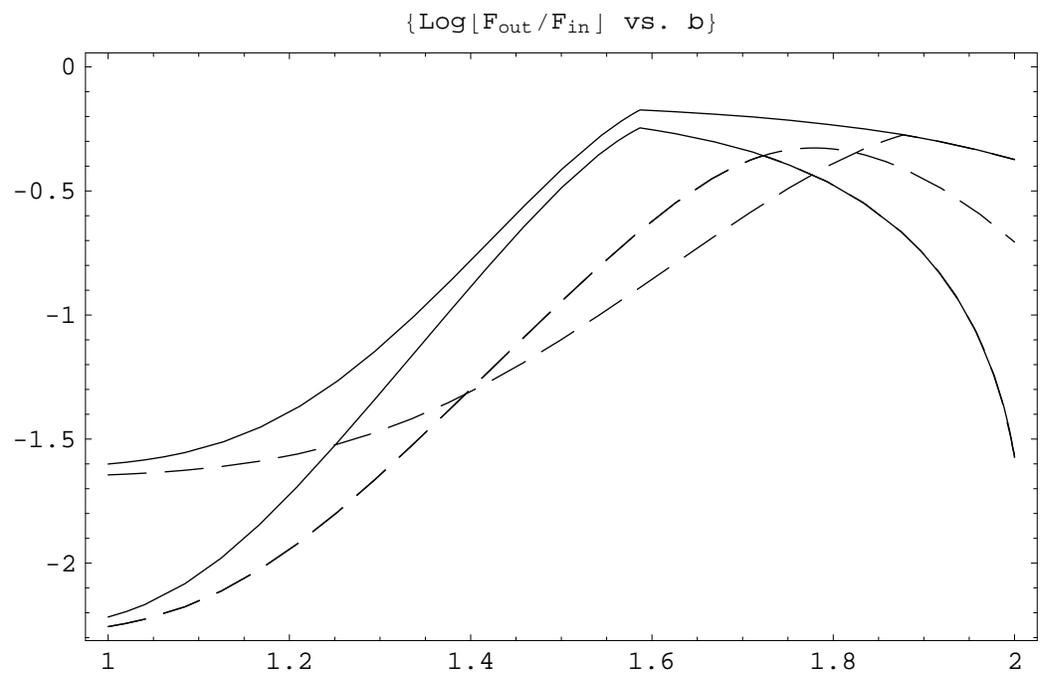
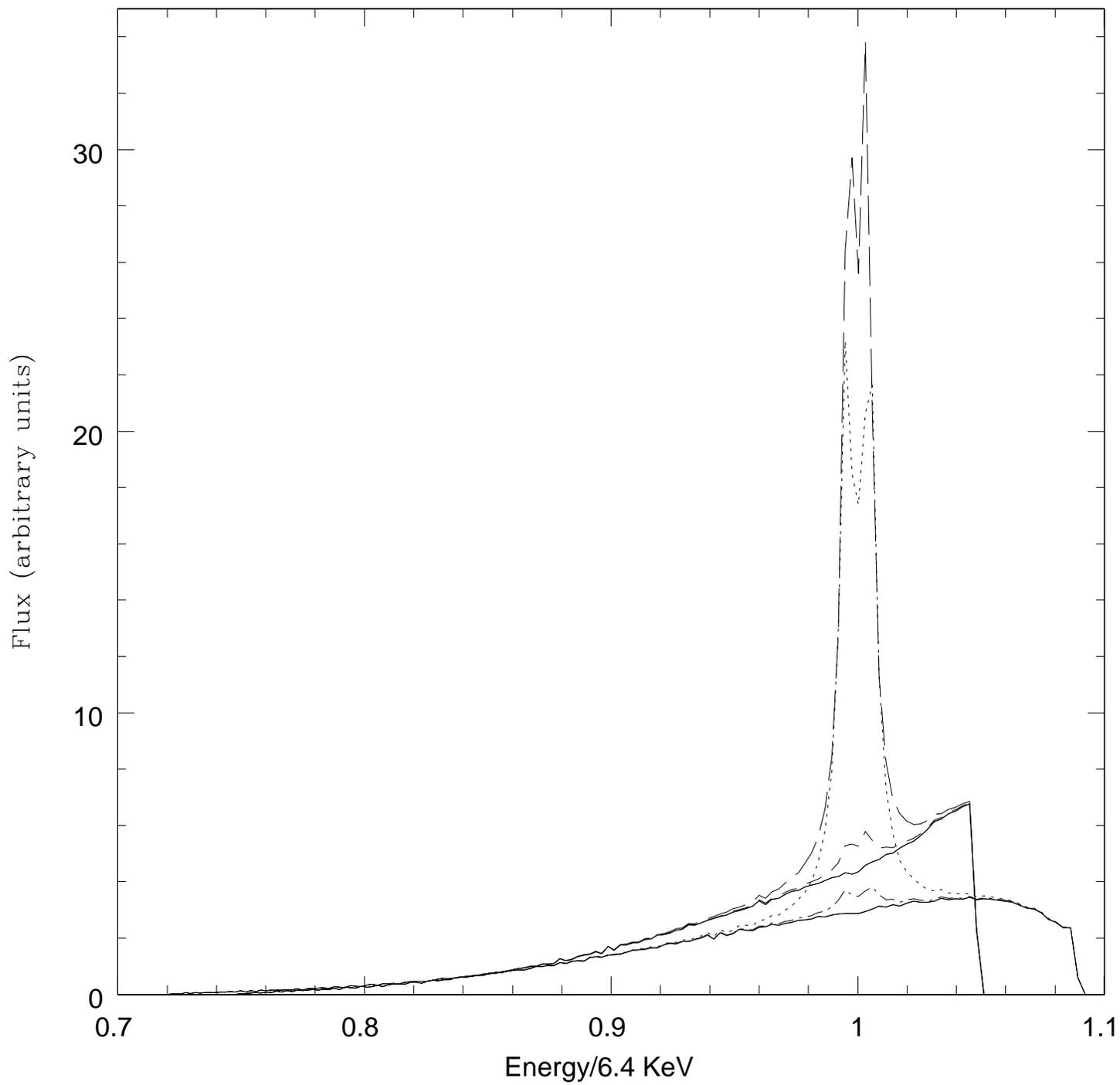
\end